\documentclass[11pt]{article}
\usepackage{amssymb}
\usepackage[hmarginratio=1:1,top=32mm,columnsep=20pt]{geometry}
\begin{document}
\begin{center}
\huge{Charged Embedded Horizons and their Area Evolution}\\[4mm]
\Large{Anushka Durg\Large{\footnote[1]{Electronic address: anushka.durg@cfrce.in}} and Aryan Bethmangalkar\Large{\footnote[2]{Electronic address: aryan.bethman@cfrce.in}} }\\[4mm]
\Large{Center for Fundamental Research and Creative Education, Bangalore, India}\\[4mm]

\end{center}
\begin{large}
\begin{abstract}
 The Embedded Horizon is defined to be a horizon that is in equilibrium with the exterior of the black hole, that is, isolated on the outside, but dynamically evolving on the inside, analogous to the inner and outer event horizons of the Reissner-Nordstrom black hole. This is shown as the result of charge variance on the horizon, which is expressed in electrical angular coordinates. The mass and energy of the black hole is discussed. The intrinsic metric is calculated  by taking the electric potential into consideration. An area law evolution law is formulated, which suggests that the inner dynamical horizon evolves until it reaches the radius of the isolated horizon. The dynamical horizon, then, reaches equilibrium with the exterior, turning into an isolated horizon, thus turning extremal.
\end{abstract}
\end{large}
\section{Introduction}
\large{Black holes are space-time structures that have been studied ever since The General Theory of Relativity has been discovered. Black holes are characterized mainly by their singularities and their horizons. In this paper, we restrict ourselves to the study of horizons. 
Horizons can be described using local or non-local geometry. Event Horizons are non-local, this implies that we need to know the entire structure of space-time to identify them, given the metric. This non-locality arises from the definition of the event horizon - the boundary of the causal past of future null infinity. But, Dynamical and Isolated Horizons are quasi-local Horizons. This means that the knowledge of the space-time in the neighborhood of such horizons is sufficient for locating them.\\ \\
It can be concluded that Dynamical and Isolated are at a greater advantage compared to the event horizon, for us to study its physics.  In this paper, we discuss Isolated and Dynamical Horizons, due to the above reasons. \\ \\
There are 3 parameters that we need to consider when we talk about black holes which are the mass, charge, and angular momentum. Physicists use these three quantities to determine the behavior and properties. In this paper, we consider a non-rotating horizon. Hence the quantity $\delta J$ is said to be zero.\\ \\
Since we have considered a non-rotating horizon(where a Dynamical horizon is embedded within an isolated horizon), the flux of gravity is said to be zero. This will be proved using Raychaudhuri’s equation [1]. From this, we deduce that the electric and magnetic components can be written using sine and cosine functions.\\ \\
In this paper, we show that charges fluctuate in section 3. We later discuss the consequences of the fluctuation of charges in section 4 and in section 5 calculate the intrinsic metric of the quasi-local space-time using the quasi-local coordinates derived in section 3.\\ \\
Much of our work is developed around Ashtekar's Isolated Horizons [1,2] and Ashtekar and Krishnan's Dynamical Horizons [2,3]. We define a new embedded horizon, from the fact that charges fluctuate. The embedded horizon is defined to be a horizon in which the dynamical horizon is embedded within the isolated horizon, much like how the inner event horizon is embedded in the outer event horizon in Reissner-Nordstrom and Kerr black holes. In section 6, we show that the inner dynamical horizon evolves until it reaches the isolated horizon, reaching equilibrium. But in order to understand these, we need to know the prerequisites explained in section 2.
\section{Preliminaries}
\large{This section consists of the prerequisites to go further in developing the above mentioned topics.}
\section*{Tetrad Formalism}
\large{In this subsection, we introduce the tetrad formalism which is a key factor in developing the mathematics of quasi-local horizons. Let us define a null tetrad $(l^a,n^a,m^a,\bar{m}^a)$ by the normalization conditions:$$l^an_a = -1$$ and $$m^a\bar{m}_a = 1$$ It must be noted that $l^a$ and $n^a$ are future-directed null normals to the horizon. The expansion of $l^a$ is given by:\begin{equation}
\theta_l = q^{ab}\nabla_al_b
\end{equation} where $q_ab$ is the intrinsic metric of the space-time. The expansion is in close relation to the area of the horizon of the black hole:\begin{equation}
\frac{dA}{dt} = \int{\theta_l \,.\, \epsilon_\Delta}
\end{equation} Given a manifold $\mathcal{M}$, we define a submanifold $\mathcal{H}$ which contains the black hole and the area it influences significantly. The induced metric is nothing but the metric on the submanifold $\mathcal{H}$. \\ \\
  \section*{ Isolated Horizons}
  \large{ An Isolated Horizon (IH) $\Delta$ is a quasi-local horizon that is defined to be in equilibrium with the exterior of the black hole. The intrinsic geometry of the horizon must remain time-independent, but the exterior can be subject to factors like radiation and other time dependent processes. An IH satisfied by the following boundary conditions (See [1] and [2]):\\ \\
   1. $\Delta$ is null and topologically $S^2 \times \mathbb{R}$ and comes equipped with a preferred foliation by 2-spheres.\\
   2. The null normal $l^a$ has vanishing expansion, $\theta_l = 0$ \\
   3. All equations of motion hold true at $\Delta$}.\\ \\
   Besides these boundary conditions, we must strictly ensure that the space-time does not allow any Killing vectors. Since this horizon is in equilibrium, this horizon must not evolve and thus, the area must remain unchanged. This is another direct effect of the time-independent case.
   
\section*{Dynamical Horizons}
\large{\textbf{Definition:} A Dynamical Horizon (DH) is smooth, three-dimensional, space-like sub-manifold of space-time which can be foliated by a family of close 2-surfaces such that on each leaf the expansion of $l^a$ vanishes and expansion of $n^a$ is negative (For more details, see [2] and [3]). \\ A DH can be time dependent and can evolve and allow interactions on it to occur. \\
\section*{The Embedded Horizon}
Now we attempt to answer as to how are IHs and DHs different from a regular event horizon. The answer is that an event horizon is non-local. In order to locate one, we need the properties of the entire structure of space-time. How we describe an event horizon today will depend on what we throw in it tomorrow [5]. But a DH or an IH is not the same. They are quasi-local. One only requires the structure of the space-time in and around the horizon, nothing more.\\ \\
One may have already noticed the similarities between IHs and DHs except one is null, and the other is space-like. Thus, we define and embedded horizon to be an isolated horizon to be a superset of the dynamical horizon, analogous to the inner and outer event horizons in the Reissner-Nordstrom solution. Thus we have imposed another condition requiring the black hole to be either charged or rotating. In this manuscript, we restrict ourselves to the charged black holes.} 
\section{Charge Fluctuations}
\large{Our aim in this section is to show that the charges on an embedded horizon tend to vary. We take into account both IHs and DHs in the below calculations. In order to do so, we use the Raychaudhuri Equation: \begin{equation}
\mathcal{L}_l \theta_l = -\frac{1}{2}\theta_l^2 - \sigma_{ab} \sigma^{ab} - R_{ab }l^a l^b
\end{equation}
where $\sigma_{ab}$ is the shear tensor and $R_{ab}$ is the Ricci Curvature Tensor. Since $\theta_l$ and $\sigma_{ab}$ vanish, we can conclude that \begin{equation}
R_{ab} l^a l^b = T_{ab} l^a l^b = 0
\end{equation}
which implies that the flux of gravity and energy-momentum is zero.\\
The flux of gravity through an isolated horizon can also be written as:\begin{equation}
\oint_{S_\Delta}\kappa \, \epsilon_\Delta
\end{equation} } where $\kappa$ is the surface gravity and $\epsilon_\Delta$ is the volume form of the horizon. From (3), (4) is zero.\\
$\kappa$ is given by:\begin{equation}
\kappa = \frac{1}{2r_\Delta} \left(1 - \frac{G(P^2_\Delta + Q^2_\Delta)}{r^2_\Delta} - \Lambda r^2_\Delta\right)
\end{equation}
where $r_\Delta$ is the radius of the horizon and $P_\Delta$ and $Q_\Delta$ are the magnetic and electric charges respectively, with $\Lambda$ as Einstein's cosmological constant. (See Reference [1])\\

From [2], the equation for gravitational flux is given by:
\begin{equation}
\mathcal{F}_g = \frac{1}{16\pi G} \int \{ |\sigma^2| +2|\zeta^2|\} \, \partial r_\Delta \, \epsilon_\Delta
\end{equation}
where $\zeta$ is the tangent vector and $\partial r_\Delta $ is the gradient of $r_\Delta$.\\ \\
Substituting (4) into (5) and equating the result to (6),
we find:
\begin{equation}
P_\Delta^2+Q_\Delta^2=0
\end{equation}
From (7), we find that :
\begin{equation}
P_\Delta=iQ_\Delta^{1/2}\,\,\,\, and \,\,\, Q_\Delta=iP_\Delta^{1/2}
\end{equation}
Therefore, they can be written in the form:
\begin{equation}
Q_\Delta=r_\Delta i sin(\alpha) \,\,\, and \,\,\, P_\Delta=r_\Delta i sin(\beta)
\end{equation} 
where $\alpha$ and $\beta$ are "angular co-ordinates" of the horizon. But in this paper, we concentrate more on electric charge as magnetically charged black holes are rather rare. 
\section{Effects of Charge Fluctuation }
\large{In this section we discuss the effects of charge variance on matter, energy and attempt to determine $\alpha$ and $\beta$ using the First Law.}\\ \\
\large{Here we consider the First law of Black Hole mechanics (For a Reissner-Nordstrom Case):
\begin{equation}
\delta M_\Delta = \frac{\kappa}{8\pi G}\delta a_\Delta +\Phi \delta Q_\Delta
\end{equation}
where $M_\Delta$ is the horizon mass, $a_\Delta$ is the surface area, $Q_\Delta$ is the charge and $\Phi$ is the potential field. From (10), we can conclude that:
\begin{equation}
\frac{\partial \kappa}{\partial Q_\Delta} = 8\pi G \, \,\frac{\partial \Phi}{\partial a_\Delta}
\end{equation}
From (4), we find that:
\begin{equation}
\Phi = \frac{1}{8\pi G} \int -\frac{GQ_\Delta}{r_\Delta^3}\,\, da_\Delta = -\frac{2Q_\Delta}{3\sqrt{\pi a_\Delta}}
\end{equation} 
with $a_\Delta = 4\pi r_\Delta^2$. Substituting this in (10) and replacing $a_\Delta$ with $4\pi r_\Delta^2$, we get the energy as:
\begin{equation}
M_\Delta = \frac{3\pi \kappa r_\Delta - GQ_\Delta^2}{6\pi G r_\Delta}
\end{equation}
This is the Energy of the horizon, which is also known as horizon mass. Solving for $Q_\Delta$, we get:
\begin{equation}
Q_\Delta = \sqrt{\frac{3\pi r_\Delta}{G}(\kappa r_\Delta^2 - 2GM_\Delta)}
\end{equation}
Our final aim is to determine the "angular co-ordinates" and we achieve  that by substituting (14) into (9) and obtain:
\begin{equation}
\alpha = sin^{-1}\left(\sqrt{-\frac{3\pi}{Gr_\Delta}(\kappa r_\Delta^2 - 2GM_\Delta)}\right)
\end{equation}\\ and
\begin{equation}
\beta = sin^{-1}\left(\sqrt{-\frac{3\pi}{Gr_\Delta}(\kappa r_\Delta^2 - 2GM_\Delta)}\pm \pi/2\right)
\end{equation}
since the electric and magnetic fields vary by a phase value of $\pi/2$  .\\ \\
\section{Intrinsic Metric}
\large{In this section, we use the co-ordinates derived in the above section and implement it into the Reissner-Nordstrom metric. Since the embedded horizon is an anologue to the Reissner-Nordstrom black hole, the metric will remain unchanged. However, we need to compute the intrinsic metric. In order to do so, we point out a cause for non-locality in the metric as well. Since the metric is dependent on the radial function, $r$ can extend infinitely, giving rise to a non-locality in the space-time. Our aim is to supress this by replacing it with a function that is inversely dependent on the radial vector. We choose the electric potential $\Phi_k$. We later define a cutoff point, wherein any particle beyond it, shall not significantly be affected by the black hole. Thus, we must acheive a spherical symmetry in the electric and magnetic angular coordinates as well. The metric should be of the form:\begin{equation}
d\tau^2 = A(Q,\Phi_k)dt^2 - B(Q,\Phi_k)dQ^2 - C(Q,\Phi_k)d\Gamma^2
\end{equation} 
where $$d\Gamma^2 = d\alpha^2 + sin^2(\alpha)d\beta^2 + d\theta^2 + sin^2(\theta)d\phi^2$$
In order to find the metric co-efficient, we consider the Reissner-Nordstrom metric: 
\begin{equation}
d\tau^2 = \left(1-\frac{2M}{r}+\frac{Q^2}{r^2}\right)dt^2-\frac{dr^2}{\left(1-2M/r+Q^2/r^2\right)}-r^2d\Omega^2
\end{equation} }
$A(Q,\Phi_k)$ is easily determined by substituting $\Phi_k$ instead of $r$. In order to determine $B(Q,\Phi_k)$ we consider writing (9) in polar coordinates:\begin{equation}
Q = e^{i\alpha}r
\end{equation}
Therefore, we can write $dr^2$ in the form of $dQ^2$ and substitute it in (18). The result is:\begin{equation}
dr^2 = e^{-2i\alpha}\left(1 - \frac{iQ^2}{\sqrt{1 + \Phi_k \Phi^k}}\right)dQ^2
\end{equation}
Thus, we can determine $B(Q,\Phi_k)$ by replacing (20) into (18).
The resulting metric is:\begin{equation}
d\tau^2 = \left(1-\frac{2M\Phi_k}{Q}+ \Phi_k \Phi^k\right)dt^2-\frac{e^{-2i\alpha}\left(1 - \frac{iQ^2}{\sqrt{1 + \Phi_k \Phi^k}}\right)dQ^2}{\left(1 - 2M\Phi_k/Q + \Phi_k \Phi^k\right)}-Q^2 e^{2i\alpha}d\Gamma^2
\end{equation}
 We now proceed to find the cutoff point mentioned earlier. In order to do so, we consider the metric written in the form:\begin{equation}
 q_{ab} = \eta_{ab} + h_{ab}
 \end{equation}
where $h_{ab}$ is a metric depending on the $r$. On observation, one finds the metric $h_{ab}$ to have the following conditions:\begin{equation}
h_{ab}(\infty) = 0 \,\,\, and \,\,\, h_{ab}(r_\Delta) = g_{ab}
\end{equation}
Thus, we can write $h_{ab}$ as:\begin{equation}
h_{ab} = \frac{r_\Delta}{r}q_{ab}
\end{equation}
To find the cutoff point, we solve the equation $$h_{ab} = 0$$ and find it to be:\begin{equation}
\Phi_{cutoff} = \frac{\pm \sqrt{M_\Delta^2 - Q_\Delta^2} - M_\Delta}{Q_\Delta}
\end{equation}
This is a point described by the potential beyond which the effects of black holes are trivial. The equation (21) can be thought of as the metric of a submanifold $\mathcal{H}$, which is the subset of $\mathcal{M}$. Thus, (21) is the intrinsic metric of $\mathcal{M}$ with the cutoff point as the intersection between the two manifolds. }\\ \\
\large{
We now discuss the gauge transforms for the fields in the space-time. We replace the standard Killing vector for symmetries with gauge freedoms from Maxwell's theory. The potential undergoes the transformation:
\begin{equation}
\Phi_k \mapsto \Phi_k - \partial_k f^k
\end{equation}
for any $f$. This transformation in the electric angular coordinates can be written as:
\begin{equation}
e^{i\alpha} \mapsto e^{i\alpha} - \partial_k f^k
\end{equation}
}
\section{Area Evolution of the Inner Dynamical Horizon}
Having calculated the intrinsic metric, we determine the area evolution of the Inner Horizon. Writing the metric in advanced and retarded co-ordinates, we have:\begin{equation}
d\tau^2 = Fdv^2 - 2HdvdQ - Q^2e^{-2i\alpha}d\Gamma^2
\end{equation}
where $v = t - Q^*$ and $dQ^* = FdQ$, $$F = \left(1-\frac{2M\Phi_k}{Q}+ \Phi_k \Phi^k\right) $$
and $$H = e^{-2i\alpha}\left(1 - \frac{iQ^2}{\sqrt{1 + \Phi_k \Phi^k}}\right)$$
The Lagrangian for radial geodesics is given by:\begin{equation}
L = F\dot{v}^2 - 2H\dot{v}\dot{Q} = 0
\end{equation}
This has a solution for $\dot{v} = 0$. The null normal $l^a$ is then given by:\begin{equation}
l^a = (1,F/2H,0,0)
\end{equation}
Then, the expansion $\theta_l$ is given by:
\begin{equation}
\theta_l = \frac{2HM\Phi_k\Phi^k}{QF}
\end{equation}
The area is then given by:\begin{equation}
A = A_-e^{\theta_lt}
\end{equation}
from (2), where $A_-$ is the area of the inner dynamical horizon initially. Solving for $\theta_l$, we find:\begin{equation}
\theta_l = \frac{ln(r^2/r_-^2)}{t}
\end{equation}
Defining a time-like vector field $\lambda$ and differentiating the expansion with respect to it, we find that:
\begin{equation}
\frac{d\theta_l}{d\lambda} = \frac{-2t'ln(r/r_-)}{t^2}
\end{equation}
where $t'$ denotes the derivative with respect to $\lambda$. This is the evolution law for the embedded horizon. Notice that at $ r = r_+$, the area stops evolving, suggesting that the DH has reached equilibrium with the exterior of the black hole and has turned into an IH. The evolution rate slowly reduces and stops at the radius of the IH, which gives the equation 
$$r_+=r_-$$
On solving that we find that the black hole has not only reached equilibrium, but also has turned extremal, i.e, its mass is equal to its charge. At this stage, the horizon does not emit any Hawking Radiation and its surface gravity $\kappa$ vanishes.
\section{Conclusion}
\large{In this article, we defined the embedded horizon and showed many physical properties it exhibits. We showed that charges fluctuate, the Hawking's Area Theorem holds true for dynamical horizons and that a dynamical horizon can become isolated, if it is given enough time. \\ \\
This supports the universal fact that all physical entities in the universe, tend to reach equilibrium and will reach equilibrium in time, including vague objects such as black holes, as this can be generalized to any horizon, including the event horizon.  \\
Another observation in section 7 is that the area does not tend to behave analogously to entropy as introduced by Bekenstein, but behaves analogously to temperature as it reaches equilibrium, since it reaches equilibrium, like temperature. Thus, there are a lot of frontiers and areas of development in this field.\\ 
}

\textbf{Acknowledgement- } Our work was carried out at Center for Fundamental Research and Creative Education (CFRCE), Bangalore, India. This paper was written under the guidance of Dr. B. S. Ramachandra, to whom we are grateful. We owe our gratitude to Vasudev  Shyam for his valuable comments and advice. We would also like to thank  our fellow researchers at CFRCE, Sagarika Rao and Rahul Balaji.\\
 }\\ \\  

\Large{\textbf{References }}\\ 
\large{

[1]  Ashtekar, A., Beetle, C., Fairhurst, S.. Mechanics of Isolated
 Horizons. Class.Quant.Grav. 17 (2000) 253-298(1999) 
\newline

[2]  Ashtekar, A.,  Krishnan, B.. Isolated and dynamical horizons and their applications. LivingRev.Rel.7 (2004) 
\newline

[3]  Ashtekar, A., Krishnan, B.. Dynamical Horizons and their properties. Phys.Rev. D68 (2003) 104030 (2003) 
 \newline
 
[4] Chandrasekhar, S. (1983). The mathematical theory of black holes. Oxford [Oxfordshire: Clarendon Press.
 \newline
 
[5] Curiel, E. The Many Definitions of a Black Hole. Nature Astronomy 2019, 3(1):27-34 (2019)}
\newline

[6] Hayward, S.A.. (2008). Dynamics of Black Holes. Advanced Science Letters. 2. 10.1166/asl.2009.1027.
\newline

[7] Hayward, S.A.. (2013). Black holes: New horizons. 10.1142/8604. 
\newline

[8]  Hawking, S,W.. (1993). Hawking on the Big Bang and Black Holes. Advanced Series in Astrophysics and Cosmology.
\newline

[9]  Traschen, J. “An introduction to black hole evaporation,” arXiv:gr-qc/0010055. }
\end{document}